\documentstyle[12pt,a4]{article}

\newcommand{\U}{{\rm U}}
\newcommand{\SU}{{\rm SU}}
\newcommand{\ol}{\overline}

\begin{document}
%%%%%%%%%%%%%%%%%%%%%%%%%%%%%%%%%%%%%%%%%%%%%%%%%%%%%%%%%%%%%%%%%%%%%%%%%%%

\begin{titlepage}
\title{\hfill\parbox{4cm}
       {\normalsize UT-06-07\\{\tt hep-th/0605097}\\May 2006}\\
       \vspace{3cm}
       Anomaly Cancellations in Brane Tilings
       \vspace{2cm}}
\author{Yosuke Imamura\thanks{E-mail: \tt imamura@hep-th.phys.s.u-tokyo.ac.jp}%
\\[20pt]
{\it Department of Physics, University of Tokyo, Tokyo 113-0033, Japan}
}
\date{}

\maketitle
\thispagestyle{empty}

\vspace{0cm}

\begin{abstract}
\normalsize
We re-interpret
the anomaly cancellation conditions
for the gauge symmetries and the baryonic flavor symmetries
in quiver gauge theories realized by the brane tilings
from the viewpoint of flux conservation on branes.
\end{abstract}

\end{titlepage}
%\tableofcontents

%%%%%%%%%%%%%%%%%%%%%%%%%%%%%%%%%%%%%%%%%%%%%%%%%%%%%%%%%%%%%%%%%%%%%%%%%%%
%%%%%%%%%%%%%%%%%%%%%%%%%%%%%%%%%%%%%%%%%%%%%%%%%%%%%%%%%%%%%%%%%%%%%%%%%%%
\section{Introduction}
The brane tilings\cite{Hanany:2005ve,Franco:2005rj,Franco:2005sm,Hanany:2005ss,Feng:2005gw,Franco:2006gc},
generalization of the brane box models\cite{Hanany:1997tb,Hanany:1998ru,Hanany:1998it},
are powerful tools to construct ${\cal N}=1$ chiral gauge theories.
They visualize the structure of gauge theories realized
on D3-branes probing toric Calabi-Yau cones as dimer diagrams on ${\bf T}^2$.
Faces, edges and vertices in the diagrams
correspond to the gauge groups, bi-fundamental matter fields
and terms in the superpotentials, respectively.
Roughly speaking, the dimer diagrams illustrate the structure of brane systems.
Each face represents a stack of D5-branes of disk topology.
If the number of the D5-branes on the face is $N$,
$\SU(N)$ gauge theory arises on the disk.
Edges in a dimer diagram
represent an NS5-brane working as a frame supporting the D5-branes on the faces.
However, the precise shape of the branes in the system
was not clear until recent.

The ``physical'' meaning of the brane tilings is
clarified in \cite{Feng:2005gw}.
Let us consider D3-branes probing a toric Calabi-Yau cone ${\cal M}$.
An ${\cal N}=1$ gauge theory is realized on the D3-branes.
The brane tiling realizing the same gauge theory is obtained
in the following way.
The toric diagram, consisting of vectors in three-dimensional integral
lattice ${\bf Z}^3$,
is often used to describe the structure of ${\cal M}$.
The lattice corresponds to the toric fiber ${\bf T}^3$ of ${\cal M}$
and the vectors represent shrinking cycles in ${\cal M}$.
Due to the Calabi-Yau condition, $c_1=0$, all the endpoints of the vectors lie
on a two-dimensional
sublattice in the ${\bf Z}^3$.
This allows us to represent the toric diagram as a two-dimensional
diagram.
Taking T-duality along the ${\bf T}^2$ specified by the two-dimensional
sublattice,
we obtain an NS5-brane wrapped on
a holomorphic curve $\Sigma$ in $({\bf C}^*)^2$\cite{Franco:2005rj}.
The curve $\Sigma$ is described by the Newton's polynomial associated with
the toric diagram.
The D3-branes in ${\cal M}$ is mapped to
D5-branes wrapped on the dual ${\bf T}^2$.
These D5-branes split into disks with
their boundaries on the NS5-brane $\Sigma$.
This system, consisting of the NS5-brane and the D5-branes, are nothing but
the brane tiling.

By performing another T-duality transformation
along a direction perpendicular to the $({\bf C}^*)^2$,
in which the NS5-brane worldvolume $\Sigma$ is embedded,
we obtain a non-compact Calabi-Yau manifold ${\cal W}$, which is mirror to ${\cal M}$.
The D5-branes are transformed to D6-branes wrapped on $3$-cycles in ${\cal W}$.
The manifold ${\cal W}$ can be represented as
${\bf C}^*$ fibration over $({\bf C}^*)^2$, and
$\Sigma$ describes the singular locus
in $({\bf C}^*)^2$ on which the ${\bf C}^*$ fiber is pinched.
In Ref. \cite{Feng:2005gw}, a simple procedure, referred to as ``untwisting'',
is proposed,
which determines the topology of $\Sigma$ and the cycles wrapped by the D6-branes
from dimer diagrams.

In this paper, we use the brane tiling picture,
while Ref. \cite{Feng:2005gw} mainly adopts the picture with
the mirror Calabi-Yau ${\cal W}$.
Because these two pictures, as mentioned in \cite{Feng:2005gw},
are directly related by T-duality,
we can use the untwisting procedure as it is to determine
the NS5-brane worldvolume $\Sigma$ and the
boundaries of D5-branes on $\Sigma$.

Let us briefly review the untwisting procedure\cite{Feng:2005gw}.
Given a dimer diagram, we first make holes on all the faces
leaving narrow region around the edges.
As a result, we have a surface
made of strips connected at the vertices.
In order to obtain the worldvolume $\Sigma$ of the NS5-brane,
we twist all the strips by $180$ degrees.
Because the graph is bipartite\cite{Franco:2005rj},
this procedure transforms the
torus with holes to another orientable surface with holes.
This surface is identified with the worldvolume $\Sigma$ of the NS5-branes.
The closed path consisting of the edges belonging to a face $a$
in the dimer diagram
becomes a zig-zag path\cite{Hanany:2005ss}
on $\Sigma$.
This path is identified with the boundary of the D5-branes
corresponding to the face $a$.

Because we now know (at least the topologies of) the shapes of branes in
brane tilings, we can use them not simply as diagrams encoding
the information of gauge theories, but as dynamical brane systems.
In this paper, we investigate the cancellation of
two kinds of anomalies from this viewpoint.
The two gauge fields on the NS5-brane,
the non-dynamical zero-form flux and the $\U(1)$ vector field,
play important roles.

In the next section,
we investigate the relation between zero-form flux conservation
on the NS5-brane and the gauge anomaly cancellation.
We show for general brane tilings that
the anomaly cancellation for the gauge symmetries
follows from the conservation of the zero-form flux
on the NS5-brane worldvolume.

In \S\ref{u1b.sec}, after reviewing a similarity of
the anomaly cancellation conditions for baryonic $\U(1)$ symmetries
with those for the gauge symmetries,
we show that a certain relation obtained from the
boundary condition for the $\U(1)$ gauge fields on NS5 and D5-branes
guarantees the cancellation of the anomaly associated with the baryonic $\U(1)$ symmetries.

%%%%%%%%%%%%%%%%%%%%%%%%%%%%%%%%%%%%%%%%%%%%%%%%%%%%%%%%%%%
\section{Gauge anomaly cancellation}
Let us consider the quiver gauge theory described by a brane tiling.
Let $n_G$ denote the number of gauge groups and $N(a)$ ($a=1,\ldots,n_G$)
be the number of colors of the $a$-th gauge group.
If they are all the same, the gauge anomalies cancel.
In this section we do not impose this condition.
Because brane tilings in general give chiral gauge theories,
we cannot take arbitrary sets of ranks.
The gauge anomaly cancellation requires
$N(a)$ satisfy certain conditions.
It is known that the number of independent ranks
of gauge groups which satisfy the anomaly cancellation conditions
is $d-2$ where $d$ is the perimeter of
the toric diagram associated with the Calabi-Yau cone ${\cal M}$
dual to the brane tiling\cite{Benvenuti:2004wx,Butti:2006hc}.
Furthermore, it is well known that
the number of gauge groups, $n_G$,
is twice the area of the toric diagram\cite{Hanany:2005ve,Franco:2005rj}.
Therefore, the number of the conditions imposed on $N(a)$ is
\begin{equation}
n_G-(d-2)=2n_{\rm internal},
\label{2internal}
\end{equation}
where $n_{\rm internal}$ is the number of internal points of the toric diagram.

What are the corresponding conditions in the brane picture?
Although
it has been anticipated
since the brane box models\cite{Hanany:1997tb,Hanany:1998ru,Hanany:1998it}
were proposed
that the conditions arise from the flux
conservation on branes,
and there are works relating
the gauge anomaly cancellation
in brane box models
to the flux conservation
and brane bendings\cite{Gimon:1998nt,Armoni:1998pf,Randall:1998hc,Leigh:1998hj,Karch:1998sj},
it has not been explicitly shown for general brane tilings
due to the lack of the precise
description of tilings
in terms of physical branes.
In this section, we clarify the relation between
the anomaly cancellation and the zero-form flux conservation
on the NS5-brane
by using the untwisting procedure proposed in \cite{Feng:2005gw}.

As we mentioned above,
each gauge group lives on a stack of D5-branes of disk topology,
and the boundary of the disk is on the NS5-brane $\Sigma$.
Let us denote the disk corresponding to the gauge group $\SU(N(a))$
by $D_a$, and
its boundary by $C_a$.
We can represent the D5-brane worldvolume as the formal sum
\begin{equation}
\sum_{a=1}^{n_G}N(a)D_a.
\end{equation}

On NS5-branes, there exist two kinds of gauge fields.
One is the $\U(1)$ vector gauge field $B$ with $2$-form flux,
and the other is the non-dynamical zero-form flux $p$.
Because the flux $p$ couples to the ``wall charge''
of boundaries of D5-branes,
the following equation holds in $\Sigma$:
\begin{equation}
\sum_{a=1}^{n_G}N(a)\delta(C_a)=dp,
\end{equation}
where $\delta(C_a)$ is the delta function with support $C_a$.
If $p\neq0$ on the NS5-brane, the brane is no longer a pure NS5-brane
but is a $(p,1)$ 5-brane.%
\footnote{The value of the flux $p$ at each puncture on $\Sigma$ represents
the D5-brane charge of the corresponding external line in the web diagram.
These numbers assigned to the external lines
are the same with the parameters $b_i$ introduced in \cite{Butti:2006hc}.}
However, we refer to it as NS5-brane in what follows for simplicity.
The curve $\Sigma$, given by Newton's polynomial associated with the toric diagram,
is a surface with some punctures.
Although it is a non-compact manifold,
the total wall charge in it must vanish for the flux to conserve
just like in the case of a compact manifold
because the punctures cannot be sources nor sinks for the
flux $p$.
This means the following homology relation in
the closure $\ol\Sigma$, which is obtained by closing the punctures in $\Sigma$, must hold:
\begin{equation}
\sum_{a=1}^{n_G}N(a)C_a=0.
\label{fluxcancel}
\end{equation}
Note that the number of independent conditions
given by (\ref{fluxcancel})
is the same with (\ref{2internal}).
This is because the genus of $\ol\Sigma$ described by the Newton's polynomial of the
toric diagram is the same with the number of the internal points
$n_{\rm internal}$ in the toric diagram,
and the number of the independent cycles on $\ol\Sigma$ is twice the genus.
The fact that (\ref{fluxcancel}) gives the same number of conditions
with those derived from the requirement of the anomaly cancellation
strongly suggests that these two sets of conditions are in fact the same.

To show this is true, we need to treat
objects in dimer diagrams.
For this purpose,
we first define some notations.
We use the following indices for faces, edges and vertices in dimer diagrams:
\begin{equation}
a,b,\ldots:\mbox{faces},\quad
I,J,\ldots:\mbox{edges},\quad
i,j,\ldots:\mbox{vertices}.
\label{indices}
\end{equation}
Because faces, edges and vertices correspond to
gauge groups, bi-fundamental matter fields
and terms in the superpotential,
respectively,
we use indices in (\ref{indices}) to label these
objects in the quiver gauge theories, too.

We use the character $\in$ to represents the adjacency.
For example, $\sum_{I\in a}$ means the summation over all the edges
adjacent to the face $a$.

If $a$ is a face and $I$ is an edge belonging to $a$,
$I\cdot a$ means the face which is adjacent to $a$ by the edge $I$.
If $b=I\cdot a$, the relation $a=I\cdot b$ also holds.

The dimer diagrams are bipartite\cite{Franco:2005rj}, namely,
all the vertices can be colored with two colors, black and white,
in such a way that
the colors of two end points of an edge are always different.
Thus we can define the orientation of edges, e.g., from white to black vertices.

In order to represent the relation of faces and edges,
we define a function
$\sigma(I,a)$.
This function, defined for a face $a$ and its side $I$,
gives $+1$ if the orientation of the edge $I$ is counter-clockwise
as a side of the polygon $a$.
Otherwise, this function gives $-1$.

We also define a function
$\sigma(a,b)$ for two adjacent faces $a$ and $b$ by
$\sigma(a,b)=\sigma(I,b)=-\sigma(I,a)$, where $I$ is the edge shared by
$a$ and $b$.
If the two faces share more than one edge,
we need to specify an edge to define this function.
This function represents the chirality of the bi-fundamental field
corresponding to the edge $I$.
If $\sigma(a,b)=+1$, the bi-fundamental field
belongs to $(\ol N(a),N(b))$ of the
$\SU(N(a))\times\SU(N(b))$, and
if $\sigma(a,b)=-1$, the field belongs to $(N(a),\ol N(b))$.

Let us represent the anomaly cancellation conditions in terms of these notations.
The bi-fundamental matter fields coupling to $\SU(N(a))$
are represented as edges $I\in a$.
If $\sigma(I,a)$ is $+1$ ($-1$),
the field belongs to the fundamental (anti-fundamental) representation
of $\SU(N(a))$, and the number of the flavours
is the size $N(I\cdot a)$ of the
gauge group for the adjacent face $I\cdot a$.
Thus, we can write the gauge anomaly cancellation conditions
as
\begin{equation}
\sum_{I\in a}\sigma(I,a)N(I\cdot a)=0,\quad
\forall a,
\label{gancan}
\end{equation}
or equivalently as
\begin{equation}
\sum_{b\in a}\sigma(a,b)N(b)=0,\quad
\forall a.
\label{gancan2}
\end{equation}
As we mentioned above, $a$ and $b$ in (\ref{gancan2}) may share
more than one edge.
In such a case, we have to sum the contribution
of all the edges shared by $a$ and $b$ separately.

In order to associate these conditions to the flux conservation
on the NS5-brane $\Sigma$, we need to re-interpret the conditions from the
viewpoint of the curve $\Sigma$.
By the untwisting procedure,
the edges belonging to a face $a$ is mapped to a zig-zag path on $\Sigma$.
This path is identified with the boundary of D5-branes, $C_a$.
An edge $I$ is always shared by two faces on ${\bf T}^2$.
This means that the edge $I$ on $\Sigma$ can be regarded as an intersection of two zig-zag paths on $\Sigma$.
Because zig-zag paths are boundaries of D5-branes,
open strings can stretch between two cycles at the intersection.
From these open strings the chiral multiplet
corresponding to the edge arises\cite{Elitzur:2000pq}.

The function $\sigma(a,b)$ is identified with the intersection number
$\langle C_a,C_b\rangle$
of the two paths $C_a$ and $C_b$ on $\Sigma$.
As we mentioned above, two faces on ${\bf T}^2$ can meet on  more than one edge.
In such a case, corresponding paths $C_a$ and $C_b$ also have more than one intersecting
point and $\sigma(a,b)$ (with $I$ implicitly specified) gives the intersection number of one of the intersecting points.
In terms of the intersection,
we can rewrite the 
condition (\ref{gancan2}) as
\begin{equation}
\langle C_a,\sum_{b=1}^{n_G} N(b)C_b\rangle=0,\quad\forall a.
\label{insercond}
\end{equation}
This relation automatically holds if the wall charge cancellation
condition (\ref{fluxcancel})
is satisfied.
The converse is also correct;
Because we know that the number of the independent conditions
is the same with the number of the independent cycles in $\Sigma$,
the condition (\ref{insercond}) is equivalent to
the condition (\ref{fluxcancel}).

%%%%%%%%%%%%%%%%%%%%%%%%%%%%%%%%%%%%%%%%%%%%%%%%%%%%%%%%%%%%%%%%%%%%%%
\section{Baryonic $\U(1)$ symmetries}\label{u1b.sec}
In this section, we consider baryonic flavor symmetries
in superconformal quiver gauge theories
with all the gauge groups having equal rank.
It is known that there is a formal similarity
between the gauge anomaly cancellation and
the baryonic $\U(1)$ anomaly cancellation\cite{Benvenuti:2004wx,Butti:2006hc}.
Because of this, we expect that the cancellation of the baryonic $\U(1)$ symmetries is also explained
by a similar mechanism associated with flux conservation on $\Sigma$.
Let us first review the similarity.

To see this similarity,
it is convenient to introduce the ``gradient'' $\beta=dN$ of the
sizes of the gauge groups $N(a)$.
(Although we assume that all the sizes $N(a)$ of the gauge groups
are equal when we consider the baryonic $\U(1)$ symmetries below,
we first consider the gauge anomalies again, which arises when
$N(a)$ depends on $a$, in order to look at the similarity
between them.)
The function $\beta$ gives the assignment of numbers for edges
and explicit definition of the gradient is given by
\begin{equation}
\beta(I)=\sum_{b\in I}\sigma(I,b)N(b).
\label{betaen1}
\end{equation}
Using the identity $\sum_{I\in a}\sigma(I,a)=0$, which holds in bipartite graphs,
we can show that the anomaly cancellation condition
(\ref{gancan})
is equivalent to
\begin{equation}
\sum_{I\in a}\beta(I)=0,\quad
\forall a.
\label{gaugecon}
\end{equation}

The reason we introduced the function $\beta(I)$ is that this function satisfies the same conditions with those
imposed on the charge assignment of the anomaly-free baryonic $\U(1)$ symmetries.
Let $Q(I)$ denote the charge assignment of a baryonic $\U(1)$ symmetry.
This must satisfy
the curl-less condition $dQ=0$, or equivalently,
\begin{equation}
\sum_{I\in i}Q(I)=0,\quad\forall i
\label{dqzero}
\end{equation}
to keep the superpotential invariant.
Any $\U(1)$ flavor symmetry must satisfy (\ref{dqzero}).
In addition to this, the baryonic symmetries must satisfy the
following condition:
\begin{equation}
\oint_\alpha Q=\oint_\beta Q=0,
\label{cacbq}
\end{equation}
where $\alpha$ and $\beta$ are paths consisting of faces and edges,
and belonging to the two independent homology classes of the torus.
The integral along a path is defined as follows:
If a closed path $\mu$ is given as a periodic sequence
$\mu=\{\ldots,I_{n-1},a_{n-1},I_n,a_n,I_{a+1},a_{n+1},\ldots\}$
whose two adjacent components (e.g. the face $a_{n-1}$ and the edge $I_n$)
are also adjacent in the dimer diagram,
the integral along $\mu$
is defined by
\begin{equation}
\oint_\mu Q\equiv \sum_n\sigma(I_n,a_n)Q(I_n).
\end{equation}
The condition (\ref{cacbq}) guarantees that
any mesonic chiral operators decouple from the symmetry.

The condition (\ref{dqzero}) and (\ref{cacbq}) are nothing but the
integrability condition which guarantees
the existence of $S(a)$ satisfying
\begin{equation}
Q=dS,
\label{qds}
\end{equation}
where the gradient $dS$ is defined in the same way with (\ref{betaen1}).
This relation is identical with the relation $\beta=dN$ for the gauge anomaly cancellation.
Furthermore, the cancellation of
the $\SU(N(a))^2\U(1)_{\rm baryonic}$ triangle anomalies
requires $Q(I)$ satisfy
\begin{equation}
\sum_{I\in a}Q(I)=0,\quad
\forall a.
\label{sumqis0}
\end{equation}
This is identical to the condition (\ref{gaugecon}) for the gauge anomaly cancellation.

From this similarity, we can conclude that
there exists almost one to one correspondence
between the baryonic $\U(1)$ symmetries and the
rank distributions of anomaly-free gauge groups.
Note that if all sizes of the gauge groups are the same,
the divergence $dN$ vanishes and
there is no corresponding baryonic symmetry.
Therefore, the number of the independent $N(a)$ is
greater than the number of the baryonic symmetries by one.
In other words, there is one-to-one correspondence between
fractional branes and anomaly-free baryonic flavor symmetries\cite{Benvenuti:2004wx,Butti:2006hc}.
It is known that the number of the baryonic $\U(1)$ symmetries
is $d-3$, where $d$ is the perimeter of the toric diagram of the dual Calabi-Yau cone ${\cal M}$\cite{Franco:2005sm}.

Using this analogy, we can immediately rewrite
the anomaly freedom condition (\ref{sumqis0})
as the following homology relation in $\ol\Sigma$:
\begin{equation}
\sum_{a=1}^{n_G}S(a)C_a=0.
\end{equation}
Formally, this is obtained by replacing $N(a)$ in (\ref{fluxcancel}) by $S(a)$.
What is the interpretation of this condition in the brane picture?
To answer this question,
we should identify the (non-dynamical) gauge fields associated with
the symmetries.

A baryon is the bound state of $N$ bi-fundamental fields.
In the brane picture,
it is a bound state of $N$ open fundamental strings
stretched between two D5-branes.
This implies that the gauge fields
couples to the baryons are the diagonal $\U(1)$ gauge
fields on D5-branes.

On the other hand, in the AdS dual picture, baryons are realized as D3-branes
wrapped on $3$-cycles in ${\cal M}$, and
by taking T-duality transformation to go to the brane picture,
the wrapped D3-branes are mapped to D-strings with their ends on the NS5-brane.
Thus, the gauge field coupling to the baryons seems to
be the $\U(1)$ gauge field
on the NS5-brane, which couples to the endpoints of D-strings.

We now have two kinds of gauge fields as candidates for the gauge fields
coupling to baryons.
In fact, the correct answer is the combination of these two.
The important fact is that the $\U(1)$ gauge fields
on the D5-branes and the NS5-brane are correlated with each other
via the boundary condition on the cycles $C_a$.
Let $A^a$ be the diagonal $\U(1)$ gauge field on $D_a$ and $B$ the
gauge field on $\Sigma$.
The boundary condition imposed on them on the boundary $C_a$ is
\begin{equation}
(A^a-\Delta B)|_{C_a}=0,
\label{boundarycons}
\end{equation}
where $\Delta B$ means the discontinuity of $B$,
and $|_{C_a}$ means the pull-back on $C_a$.
An easy way to obtain this boundary condition is
to transform the system to an NS5-D4 system by T-duality,
and lift it to
a smooth M5-brane
configuration.
The boundary condition (\ref{boundarycons})
is obtained from the continuity of the two-form gauge field on the M5-brane.
If we denote the four-dimensional (non-dynamical) gauge field
for a baryonic symmetry by ${\cal A}$,
the gauge fields on branes are decomposed as
\begin{equation}
A^a=s^a{\cal A},\quad
B=f{\cal A}
\label{decomposition}
\end{equation}
where $s^a$ are function on the disks $D^a$,
and $f$ is a function on the curve $\Sigma$.
These functions depend only on the coordinates of the internal space,
while ${\cal A}$ is a $1$-form depending on the coordinates in the four-dimensional spacetime.
For the symmetry not to be broken, the gauge field ${\cal A}$
must be massless, and this implies that
the functions $s^a$ and $f$ must be constant except on the boundaries $C_a$.
The constants $s^a$ assigned to $D_a$ should be identified with
the function $S(a)$ in (\ref{qds}).
Because a bi-fundamental field arises from
open strings at the intersection of two cycles $C_a$ and $C_b$ on $\Sigma$,
its charge for the baryonic $\U(1)$ symmetry
is given as the difference of $S(a)$ and $S(b)$.
This is the physical interpretation of the relation (\ref{qds}).
By substituting (\ref{decomposition}) into (\ref{boundarycons}),
we obtain the following equation for the function $f$:
\begin{equation}
\sum_{a=1}^{n_G}S(a)\delta(C_a)=df.
\end{equation}
Thus we can treat $S(a)$ as wall charges carried by $C_a$,
and $f$ as a zero-form flux coupling to the charges $S(a)$.
This relation is formally identical to that between the wall charges
of D5-brane boundaries
and
the zero form flux $p$ discussed in the previous section.
For the flux conservation,
the total charge must vanish;
\begin{equation}
\sum_{a=1}^{n_G} S(a)C_a=0.
\end{equation}
Just like in the case of the gauge anomalies,
this condition guarantees the cancellation of the
anomalies of the baryonic $\U(1)$ symmetries.

%%%%%%%%%%%%%%%%%%%%%%%%%%%%%%%%%%%%%%%%%%%%%%%%%%%%%%%%%%
\section{Conclusions}
By using the untwisted procedure proposed in \cite{Feng:2005gw},
which relates a dimer diagram with the holomorphic curve $\Sigma$
representing the NS5-brane in the corresponding brane tiling,
we showed that the anomaly cancellations of the gauge symmetries and the
baryonic flavor symmetries
are explained by considering the behavior of two kinds of gauge fields
on the NS5-brane.

The zero-form flux conservation on the NS5-brane
requires cancellation of the wall charge
of the boundaries of the D5-branes.
This imposes conditions on the numbers of D5-branes
on the faces of the dimer diagram, and
these conditions guarantee the gauge anomaly cancellation.

We identified the gauge fields coupling to the
baryonic $\U(1)$ symmetries with combinations
of the vector fields on the D5-branes and the NS5-brane
in the brane tiling.
The correlation of these vector fields
at the junctions of the NS5 and D5-branes
imposes conditions on the baryonic $\U(1)$ charge assignment for the
bi-fundamental fields.
These conditions guarantee the cancellation
of the anomaly of the baryonic $\U(1)$ symmetries.

\section*{Acknowledgements}
I would like to thank
Y.~Nakayama,
Y.~Tachikawa,
and M.~Yamazaki,
for valuable discussions.
This work is supported in part by
a Grant-in-Aid for Scientific Research
(\#17540238)
from the Japan Ministry of Education, Culture, Sports,
Science and Technology.

\end{document}